\documentclass{article}

\usepackage{amsmath}
\usepackage{graphicx}
\usepackage{color}
\usepackage{url}
\makeatletter
\newcommand{\bm}[1]{{\mbox{\boldmath $#1$}}}
\let\saved@includegraphics\includegraphics
\AtBeginDocument{\let\includegraphics\saved@includegraphics}
\renewenvironment*{figure}{\@float{figure}}{\end@float}

\newcommand{\mnras}{Mon. Not. R. Astron. Soc.}

\newcommand{\apj}{Astrophys. J.}
\newcommand{\apjl}{Astrophys. J.}

\newcommand{\aap}{Astron. Astrophys}

\title{Solar differential rotation reproduced with high-resolution simulation}

\author
{H. Hotta$^{1\ast}$, K. Kusano$^{2}$\\
\\
\normalsize{$^{1}$Department of Physics, Graduate School of Science, Chiba University,}\\
\normalsize{1-33 Yayoi-cho, Inage-ku, Chiba 263-8522, Japan}\\
\normalsize{$^{2}$Institute for Space-Earth Environmental Research, }\\
\normalsize{Nagoya University, Furo-cho, Chikusa-ku, Nagoya, Aichi 464-8601, Japan}
\\
\\
\normalsize{$^\ast$To whom correspondence should be addressed; E-mail:  hotta@chiba-u.jp.}
}

\date{}

\begin{document}

\baselineskip24pt


\maketitle


  {\bf
  The Sun rotates differentially with a fast equator and slow pole\cite{1998ApJ...505..390S}.
  Convection in the solar interior is thought to maintain the differential rotation. However, although many numerical simulations have been conducted to reproduce the solar differential rotation\cite{2000ApJ...532..593M,2008ApJ...689.1354B,2013ApJ...762...73N,2015ApJ...798...51H,2016Sci....351..1427,2018ApJ...860L..24H}, previous high-resolution calculations with solar parameters fall into the anti-solar (fast pole) differential rotation regime. Consequently, we still do not know the true reason why the Sun has a fast-rotating equator. While the construction of the fast equator requires a strong rotational influence on the convection, the previous calculations have not been able to achieve the situation without any manipulations. The problem is called convective conundrum\cite{2016AdSpR..58.1475O}. The convection and the differential rotation in numerical simulations were different from the observations. Here, we show that a high-resolution calculation succeeds in reproducing the solar-like differential rotation. Our calculations indicate that the strong magnetic field generated by a small-scale dynamo has a significant impact on thermal convection. 
  The successful reproduction of the differential rotation, convection, and magnetic field achieved in our calculation is an essential step to understanding the cause of the most basic nature of solar activity, specifically, the 11-year cycle of sunspot activity.
  }

In this study, we drastically increase the resolution using the Supercomputer Fugaku to investigate the possible influence of the magnetic field on the differential rotation. We perform three cases, Low, Middle, and High, where the numbers of grid points are $(N_r,N_\theta,N_\phi,N_\mathrm{YY})=(96,384,1156,2)$, $(192,768,2312,2)$, and $(384,1536,4608,2)$, respectively. $N_r$, $N_\theta$, and $N_\phi$ are the radial, latitudinal, and longitudinal grid points, respectively. $N_{YY}$ is the factor from the Yin-Yang grid\cite{2004GGG.....5.9005K}. In the ordinary spherical grid, the numbers of grid points are $(N_r,N_\theta,N_\phi)=(96,768,1536)$, $(192,1536,3072)$, and $(384,3072,6144)$ in the Low, Middle, and High cases, respectively. Note that the resolution is fairly high even in the Low case compared with previous studies. We adopt solar stratification\cite{1996Sci...272.1286C}, solar rotation, and solar luminosity, and exclude any type of explicit diffusivity to maintain high resolution. Details of the numerical method are found in the Method section. We continue these calculations for 4000 days. The temporal evolutions of the energies are shown in Supplementary Figure 1.\par
Figure \ref{fig:overall}A and B shows three-dimensional volume rendering of the normalized entropy and the magnetic field strength in the High case, respectively. 
The maximum magnetic field strength exceeds 80 kG, which is a significant superequipartition magnetic field.

\begin{figure}[htbp]
  \begin{center}
      \includegraphics[width=\textwidth]{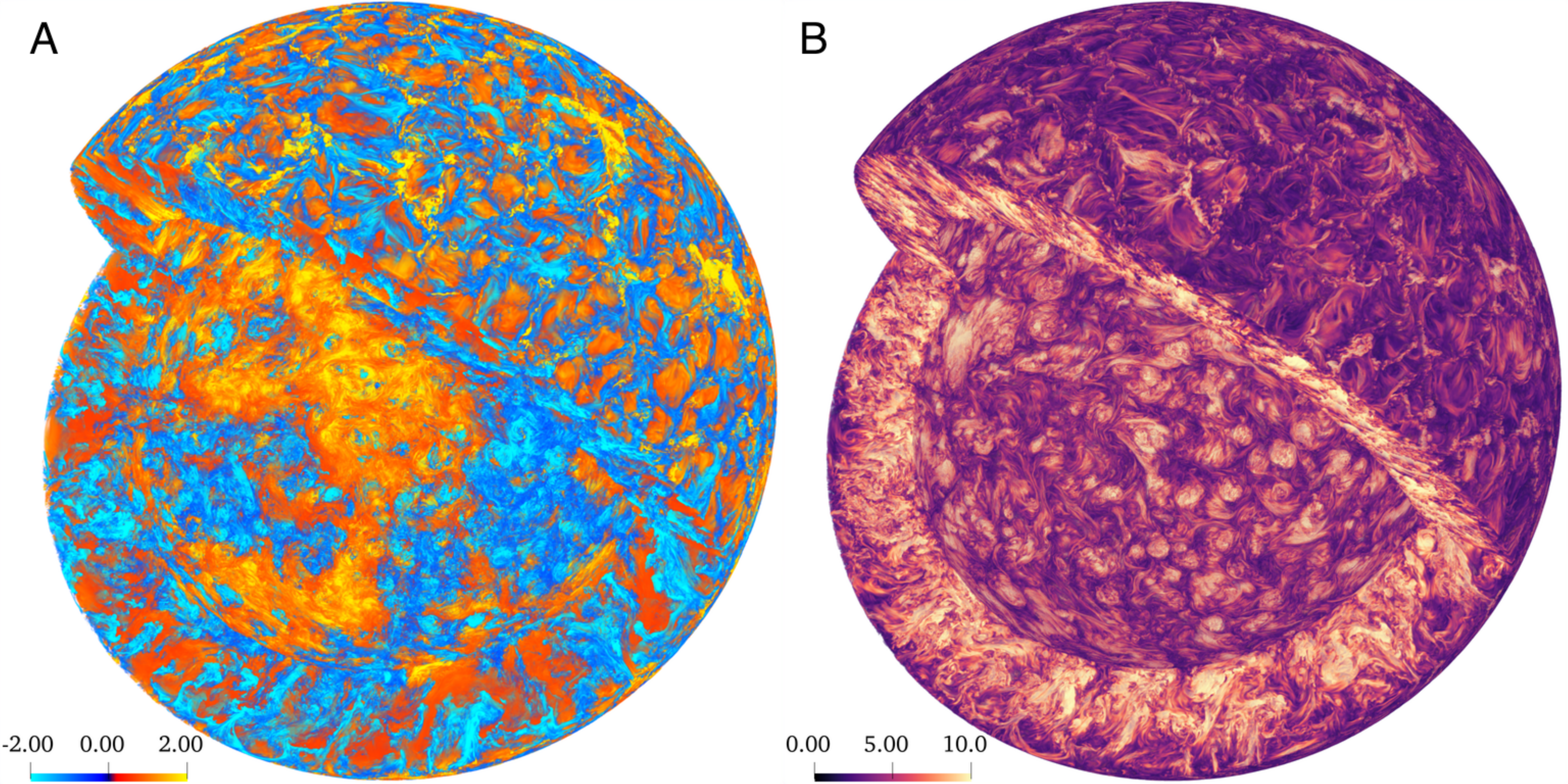}
      \caption{{\bf Overall structure of convection and magnetic field.} Panels A and B show the normalized entropy and the magnetic field strength respectively. The magnetic field strength is shown in the unit of kG. A quadrant of sphere is striped to see the interior of the Sun.}
      \label{fig:overall}
  \end{center}
\end{figure}
\par
\begin{figure}[htbp]
  \begin{center}
      \includegraphics[width=0.95\textwidth]{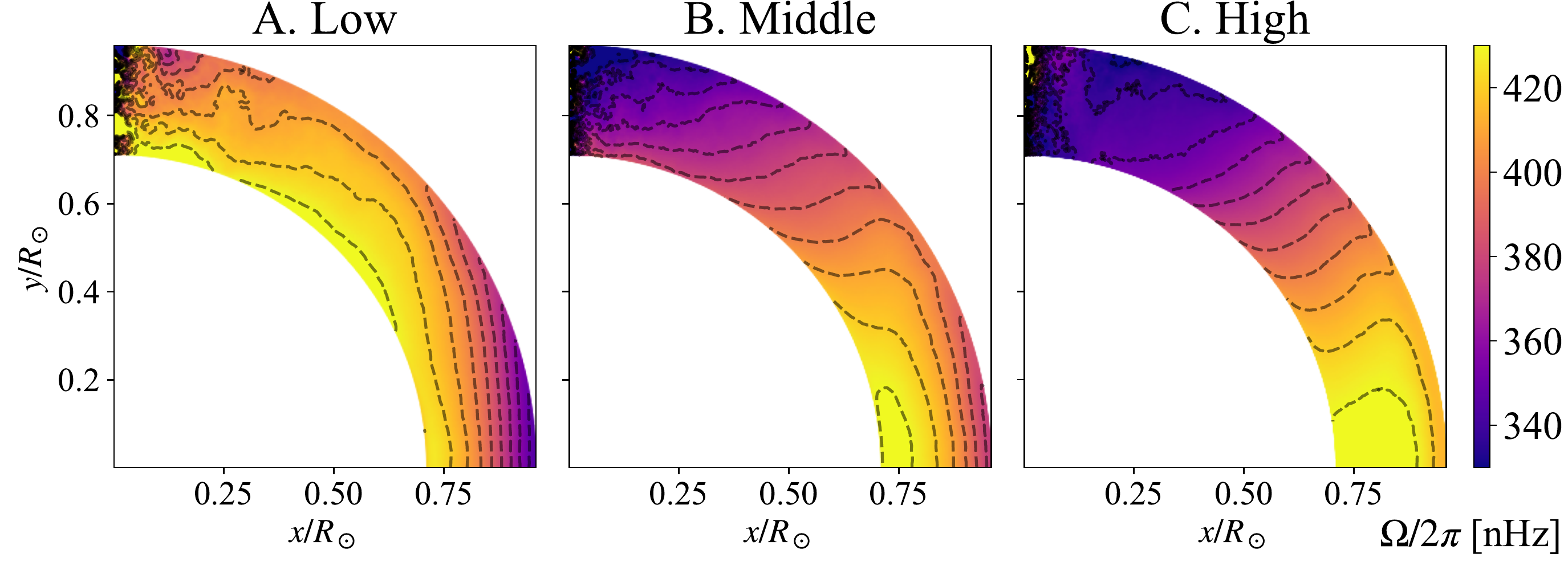}
      \caption{{\bf Dependence of differential rotation on the resolution.} Panels A, B, and C show the results from the Low, Middle, and High cases, respectively. The value $\Omega/2\pi$ is shown in units of nHz, where $\Omega$ is the angular velocity. The dashed lines show the values from 330 to 430 nHz in 10 nHz increments. The results in the northern and southern hemispheres are averaged.}
      \label{fig:differential_rotation}
  \end{center}
\end{figure}

Figure \ref{fig:differential_rotation} shows the dependence of the differential rotation on the resolution. Panels A, B, and C show the results for the Low, Middle, and High cases, respectively. In the Low case (Figure \ref{fig:differential_rotation}A), we obtain the fast pole and slow equator, which is consistent with previous studies. In the Middle case (Figure \ref{fig:differential_rotation}B), the differential rotation becomes more solar-like, while we still see a significant decrease of the angular velocity in the near-surface region around the equator. Even in the Middle case, the resolution is high as a long-term full spherical dynamo calculation. In the High case (Figure \ref{fig:differential_rotation}C), we nicely reproduce the solar-like differential rotation, specifically, the equator region is rotating faster than the pole. Similar to the real Sun, the topology of the differential rotation deviates from the Taylor--Proudman-like profile, i.e. the contour lines of our differential rotation are not aligned to the rotational axis. 
We do not force the entropy gradient at the bottom boundary\cite{2006ApJ...641..618M}, but the efficient small-scale dynamo increases the latitudinal entropy gradient as found by Hotta (2018)\cite{2018ApJ...860L..24H}. A differential rotation in a hydrodynamic calculation without the magnetic field is also shown in Supplementary Figure 3.

\par
Figure \ref{fig:velocity_magnetic_field} shows the convection and magnetic field properties. Panel A shows root-mean-square (RMS) velocity $v_\mathrm{RMS}$. Higher resolution tends to show a smaller amplitude of the convection. 
The decrease in the convection velocity reduces the Rossby number. Figure \ref{fig:velocity_magnetic_field}B shows the RMS magnetic field ($B_\mathrm{RMS}$: solid line) and equipartition magnetic field ($B_\mathrm{eq}$: dotted line), where $B_\mathrm{eq}=\sqrt{4\pi\rho_0 v_\mathrm{RMS}}$ and $\rho_0$ is the background density. The RMS magnetic fields monotonically increase with the resolution. In the Low case, the magnetic field is always smaller than the equipartition magnetic field. In previous studies, the RMS magnetic field reaches 10--20\% of the equipartition magnetic field\cite{2014ApJ...789...35F}. The system reaches an efficient small-scale dynamo regime even in the Low case because the magnetic field achieves an almost equipartition level. In the Middle case, the superequipartition magnetic field ($B_\mathrm{RMS}>B_\mathrm{eq}$) is reproduced in the bottom half of the convection zone. 
In the High case, the magnetic energy exceeds the kinetic energy in all layers in the convection zone. This strong magnetic field suppresses the convection velocity significantly. {In the High case, the stretching becomes weaken and the compression increases. The generation mechanism of the magnetic field is discussed also in Supplementary Figures 4 and 5.

\begin{figure}[htbp]
  \begin{center}
      \includegraphics[width=0.9\textwidth]{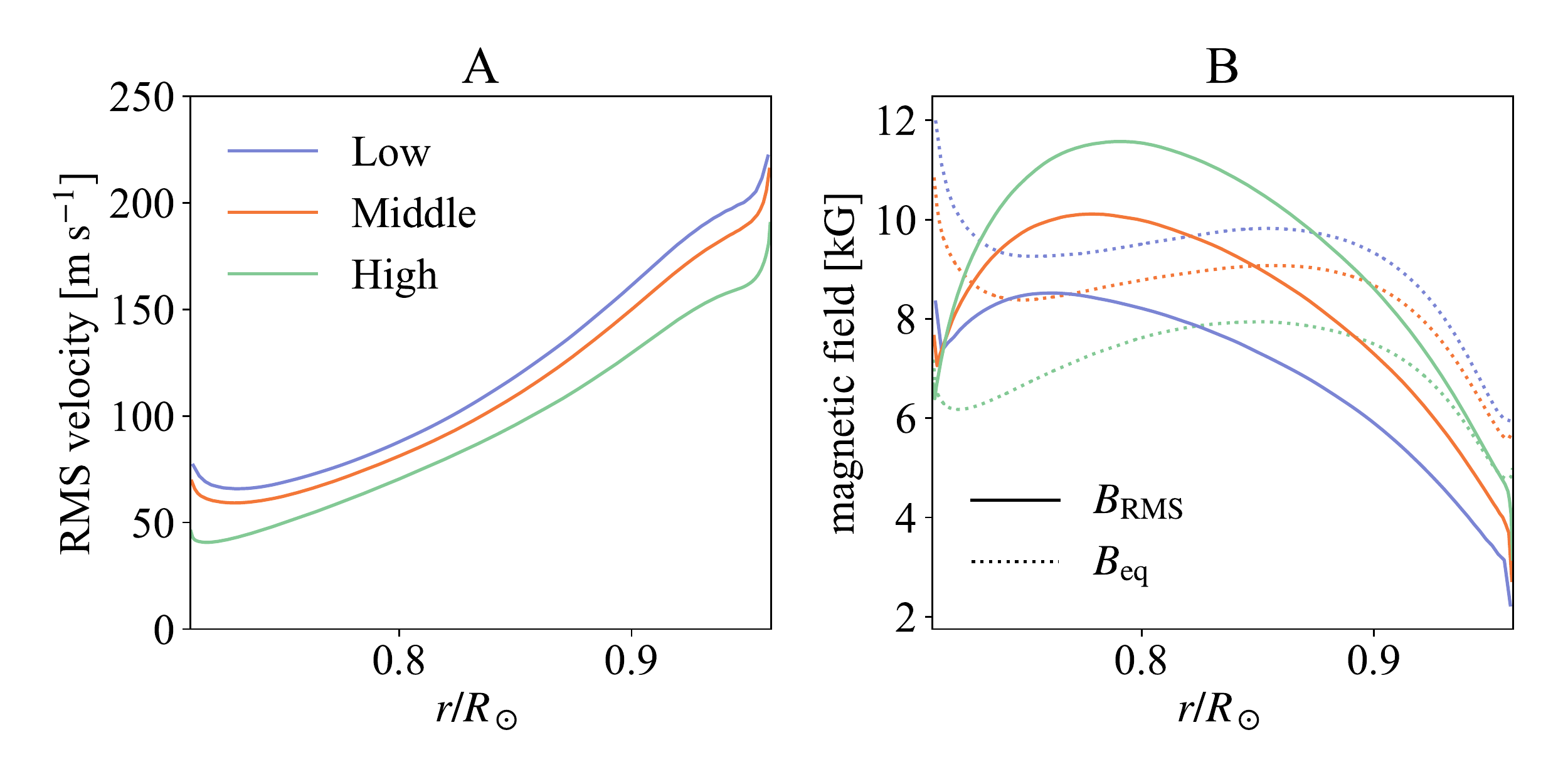}
      \caption{{\bf Dependence of the convection and magnetic field properties on the resolution.} Panel A shows the RMS velocity, and panel B shows the RMS (solid lines) and equipartition (dotted lines) magnetic field strengths. The blue, orange, and green lines show the results from the Low, Middle, and High cases, respectively.}
      \label{fig:velocity_magnetic_field}
  \end{center}
\end{figure}

\par
Figure \ref{fig:spectra} shows the kinetic (solid line) and magnetic (dotted line) energy spectra at $r=0.83R_\odot$. In the Low case, the magnetic energy exceeds the kinetic energy only on a small scale ($\ell>100$). This is a clear sign of the efficient small-scale dynamo\cite{2015ApJ...803...42H}. In the Middle case, the turnover scale of the superequipartition magnetic field moves to a larger scale ($\ell\sim45$). While in the small scale ($\ell>10$), the kinetic energy in the Middle case is smaller than that of the Low case because of stronger Lorentz force feedback, the kinetic energy does not change in the large-scale ($\ell<10$). In the High case, the magnetic energy exceeds the kinetic energy on almost all the scales. The kinetic energy is also reduced in all the scales. Because of this kinetic energy suppression in the High case, the peak of the kinetic energy is shifted from $\ell\sim6$ (Low and Middle cases) to $\ell\sim30$. These spectra variations indicate that the dynamo in the High case is qualitatively different from the others.

\begin{figure}[htbp]
  \begin{center}
      \includegraphics[width=0.6\textwidth]{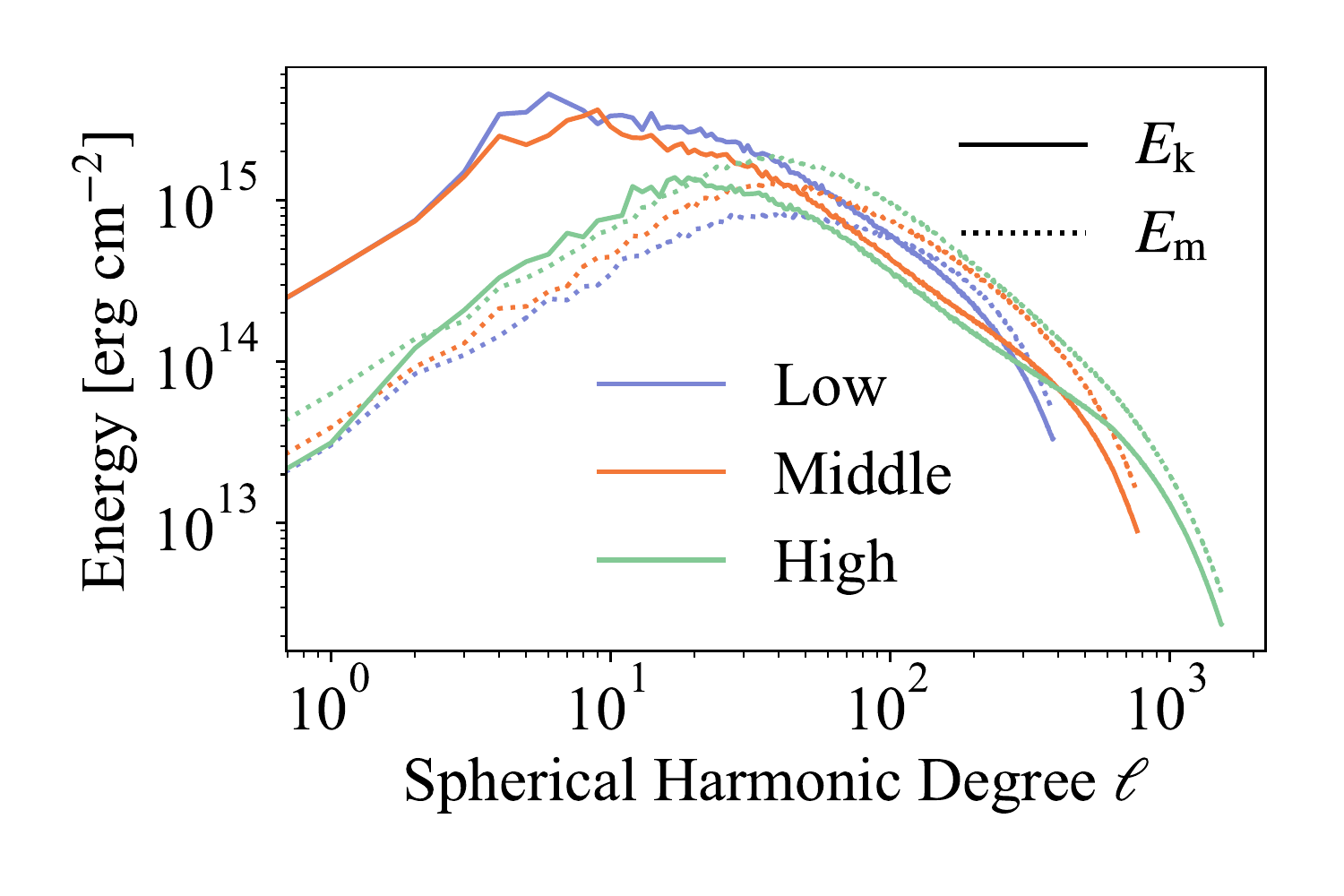}
      \caption{{\bf Dependence of kinetic and magnetic energy spectra on the resolution.} The spectra at $r=0.83R_\odot$ are shown. The solid and dotted lines show the kinetic and magnetic energy spectra, respectively. The blue, orange, and green lines show the spectra in Low, Middle, and High cases, respectively. In this plot, only the $m\neq 0$ mode is shown to exclude contributions from the differential rotation, where $m$ is the spherical harmonic order.}
      \label{fig:spectra}
  \end{center}
\end{figure}

\section*{Discussion}

In this study, we reproduce the solar-like differential rotation in an high-resolution calculation with solar parameters, such as the stratification, the luminosity, and the rotation rate. Our calculation results show that the Sun has a marginal Rossby number between anti-solar (fast pole) and solar-like (fast equator) differential rotation, and we need a sophisticated treatment of the thermal convection and the magnetic field to reproduce the differential rotation. 
We offer a path forward in resolving the convective conundrums. Significant reduction in the convective energy in the large-scale ($\ell<30$) is a promising trend to solve the other part of the convective conundrum, i.e. the energy spectra obtained by helioseismology\cite{2012PNAS..10911928H}. Because the observational result for the energy spectra is still controversial, detailed comparisons between numerical simulations and observations are needed to solve the problem. Furthermore, by appropriately reproducing the differential rotation, the convection will lead to a correct understanding of the generation of the large-scale magnetic field and cycle. In this study, the dynamo has not constructed the large-scale magnetic field (see Supplementary Figure 2) probably due to lack of the calculation time for the large-scale dynamo or the insufficient suppression of the convection velocity.
It is possible that the longer time calculation changes the amplitude of the magnetic field. This is still out of our reach. In addition, we have not reached the numerical convergence where the result does not change with doubling the resolution. We cannot rule out the further change of the differential rotation in higher-resolution simulations. We expect the further resolution leads to a stronger magnetic field which more suppresses the convection velocity. This is a good factor for the construction of the large-scale magnetic field as well. The higher resolution simulation is still on demand.


\begin{thebibliography}{10}
    \expandafter\ifx\csname url\endcsname\relax
      \def\url#1{\texttt{#1}}\fi
    \expandafter\ifx\csname urlprefix\endcsname\relax\def\urlprefix{URL }\fi
    \providecommand{\bibinfo}[2]{#2}
    \providecommand{\eprint}[2][]{\url{#2}}
    
    \bibitem{1998ApJ...505..390S}
    \bibinfo{author}{{Schou}, J.} \emph{et~al.}
    \newblock \bibinfo{title}{{Helioseismic Studies of Differential Rotation in the
      Solar Envelope by the Solar Oscillations Investigation Using the Michelson
      Doppler Imager}}.
    \newblock \emph{\bibinfo{journal}{\apj}} \textbf{\bibinfo{volume}{505}},
      \bibinfo{pages}{390--417} (\bibinfo{year}{1998}).
    
    \bibitem{2000ApJ...532..593M}
    \bibinfo{author}{{Miesch}, M.~S.} \emph{et~al.}
    \newblock \bibinfo{title}{{Three-dimensional Spherical Simulations of Solar
      Convection. I. Differential Rotation and Pattern Evolution Achieved with
      Laminar and Turbulent States}}.
    \newblock \emph{\bibinfo{journal}{\apj}} \textbf{\bibinfo{volume}{532}},
      \bibinfo{pages}{593--615} (\bibinfo{year}{2000}).
    
    \bibitem{2008ApJ...689.1354B}
    \bibinfo{author}{{Brown}, B.~P.}, \bibinfo{author}{{Browning}, M.~K.},
      \bibinfo{author}{{Brun}, A.~S.}, \bibinfo{author}{{Miesch}, M.~S.} \&
      \bibinfo{author}{{Toomre}, J.}
    \newblock \bibinfo{title}{{Rapidly Rotating Suns and Active Nests of
      Convection}}.
    \newblock \emph{\bibinfo{journal}{\apj}} \textbf{\bibinfo{volume}{689}},
      \bibinfo{pages}{1354--1372} (\bibinfo{year}{2008}).
    \newblock 
    
    \bibitem{2013ApJ...762...73N}
    \bibinfo{author}{{Nelson}, N.~J.}, \bibinfo{author}{{Brown}, B.~P.},
      \bibinfo{author}{{Brun}, A.~S.}, \bibinfo{author}{{Miesch}, M.~S.} \&
      \bibinfo{author}{{Toomre}, J.}
    \newblock \bibinfo{title}{{Magnetic Wreaths and Cycles in Convective Dynamos}}.
    \newblock \emph{\bibinfo{journal}{\apj}} \textbf{\bibinfo{volume}{762}},
      \bibinfo{pages}{73} (\bibinfo{year}{2013}).
    \newblock 
    
    \bibitem{2015ApJ...798...51H}
    \bibinfo{author}{{Hotta}, H.}, \bibinfo{author}{{Rempel}, M.} \&
      \bibinfo{author}{{Yokoyama}, T.}
    \newblock \bibinfo{title}{{High-resolution Calculation of the Solar Global
      Convection with the Reduced Speed of Sound Technique. II. Near Surface Shear
      Layer with the Rotation}}.
    \newblock \emph{\bibinfo{journal}{\apj}} \textbf{\bibinfo{volume}{798}},
      \bibinfo{pages}{51} (\bibinfo{year}{2015}).
    \newblock 
    
    \bibitem{2016Sci....351..1427}
    \bibinfo{author}{{Hotta}, H.}, \bibinfo{author}{{Rempel}, M.} \&
      \bibinfo{author}{{Yokoyama}, T.}
    \newblock \bibinfo{title}{{Large-scale magnetic fields at high Reynolds numbers
      in magnetohydrodynamic simulations}}.
    \newblock \emph{\bibinfo{journal}{Science}} \textbf{\bibinfo{volume}{351}},
      \bibinfo{pages}{1427--1430} (\bibinfo{year}{2016}).
    
    \bibitem{2018ApJ...860L..24H}
    \bibinfo{author}{{Hotta}, H.}
    \newblock \bibinfo{title}{{Breaking Taylor-Proudman Balance by Magnetic Fields
      in Stellar Convection Zones}}.
    \newblock \emph{\bibinfo{journal}{\apjl}} \textbf{\bibinfo{volume}{860}},
      \bibinfo{pages}{L24} (\bibinfo{year}{2018}).
    \newblock 
    
    \bibitem{2016AdSpR..58.1475O}
    \bibinfo{author}{{O'Mara}, B.}, \bibinfo{author}{{Miesch}, M.~S.},
      \bibinfo{author}{{Featherstone}, N.~A.} \& \bibinfo{author}{{Augustson},
      K.~C.}
    \newblock \bibinfo{title}{{Velocity amplitudes in global convection
      simulations: The role of the Prandtl number and near-surface driving}}.
    \newblock \emph{\bibinfo{journal}{Advances in Space Research}}
      \textbf{\bibinfo{volume}{58}}, \bibinfo{pages}{1475--1489}
      (\bibinfo{year}{2016}).
    \newblock 
    
    \bibitem{2004GGG.....5.9005K}
    \bibinfo{author}{{Kageyama}, A.} \& \bibinfo{author}{{Sato}, T.}
    \newblock \bibinfo{title}{{``Yin-Yang grid'': An overset grid in spherical
      geometry}}.
    \newblock \emph{\bibinfo{journal}{Geochemistry, Geophysics, Geosystems}}
      \textbf{\bibinfo{volume}{5}}, \bibinfo{pages}{Q09005} (\bibinfo{year}{2004}).
    \newblock 
    
    \bibitem{1996Sci...272.1286C}
    \bibinfo{author}{{Christensen-Dalsgaard}, J.} \emph{et~al.}
    \newblock \bibinfo{title}{{The Current State of Solar Modeling}}.
    \newblock \emph{\bibinfo{journal}{Science}} \textbf{\bibinfo{volume}{272}},
      \bibinfo{pages}{1286--1292} (\bibinfo{year}{1996}).
    
    \bibitem{2006ApJ...641..618M}
    \bibinfo{author}{{Miesch}, M.~S.}, \bibinfo{author}{{Brun}, A.~S.} \&
      \bibinfo{author}{{Toomre}, J.}
    \newblock \bibinfo{title}{{Solar Differential Rotation Influenced by
      Latitudinal Entropy Variations in the Tachocline}}.
    \newblock \emph{\bibinfo{journal}{\apj}} \textbf{\bibinfo{volume}{641}},
      \bibinfo{pages}{618--625} (\bibinfo{year}{2006}).
    
    \bibitem{2014ApJ...789...35F}
    \bibinfo{author}{{Fan}, Y.} \& \bibinfo{author}{{Fang}, F.}
    \newblock \bibinfo{title}{{A Simulation of Convective Dynamo in the Solar
      Convective Envelope: Maintenance of the Solar-like Differential Rotation and
      Emerging Flux}}.
    \newblock \emph{\bibinfo{journal}{\apj}} \textbf{\bibinfo{volume}{789}},
      \bibinfo{pages}{35} (\bibinfo{year}{2014}).
    \newblock 
    
    \bibitem{2015ApJ...803...42H}
    \bibinfo{author}{{Hotta}, H.}, \bibinfo{author}{{Rempel}, M.} \&
      \bibinfo{author}{{Yokoyama}, T.}
    \newblock \bibinfo{title}{{Efficient Small-scale Dynamo in the Solar Convection
      Zone}}.
    \newblock \emph{\bibinfo{journal}{\apj}} \textbf{\bibinfo{volume}{803}},
      \bibinfo{pages}{42} (\bibinfo{year}{2015}).
    \newblock 
    
    \bibitem{2012PNAS..10911928H}
    \bibinfo{author}{{Hanasoge}, S.~M.}, \bibinfo{author}{{Duvall}, T.~L.} \&
      \bibinfo{author}{{Sreenivasan}, K.~R.}
    \newblock \bibinfo{title}{{Anomalously weak solar convection}}.
    \newblock \emph{\bibinfo{journal}{Proceedings of the National Academy of
      Science}} \textbf{\bibinfo{volume}{109}}, \bibinfo{pages}{11928--11932}
      (\bibinfo{year}{2012}).
    \newblock 
    
    \bibitem{2014ApJ...786...24H}
    \bibinfo{author}{{Hotta}, H.}, \bibinfo{author}{{Rempel}, M.} \&
      \bibinfo{author}{{Yokoyama}, T.}
    \newblock \bibinfo{title}{{High-resolution Calculations of the Solar Global
      Convection with the Reduced Speed of Sound Technique. I. The Structure of the
      Convection and the Magnetic Field without the Rotation}}.
    \newblock \emph{\bibinfo{journal}{\apj}} \textbf{\bibinfo{volume}{786}},
      \bibinfo{pages}{24} (\bibinfo{year}{2014}).
    \newblock 
    
    \bibitem{2019SciA....eaau2307}
    \bibinfo{author}{{Hotta}, H.}, \bibinfo{author}{{Iijima}, H.} \&
      \bibinfo{author}{{Kusano}, K.}
    \newblock \bibinfo{title}{{Weak influence of near-surface layer on solar deep
      convection zone revealed by comprehensive simulation from base to surface}}.
    \newblock \emph{\bibinfo{journal}{Science Advances}}
      \textbf{\bibinfo{volume}{5}}, \bibinfo{pages}{eaau2307}
      (\bibinfo{year}{2019}).
    
    \bibitem{2020MNRAS.494.2523H}
    \bibinfo{author}{{Hotta}, H.} \& \bibinfo{author}{{Iijima}, H.}
    \newblock \bibinfo{title}{{On rising magnetic flux tube and formation of
      sunspots in a deep domain}}.
    \newblock \emph{\bibinfo{journal}{\mnras}} \textbf{\bibinfo{volume}{494}},
      \bibinfo{pages}{2523--2537} (\bibinfo{year}{2020}).
    \newblock 
    
    \bibitem{2012A&A...539A..30H}
    \bibinfo{author}{{Hotta}, H.}, \bibinfo{author}{{Rempel}, M.},
      \bibinfo{author}{{Yokoyama}, T.}, \bibinfo{author}{{Iida}, Y.} \&
      \bibinfo{author}{{Fan}, Y.}
    \newblock \bibinfo{title}{{Numerical calculation of convection with reduced
      speed of sound technique}}.
    \newblock \emph{\bibinfo{journal}{\aap}} \textbf{\bibinfo{volume}{539}},
      \bibinfo{pages}{A30} (\bibinfo{year}{2012}).
    \newblock 
    
    \bibitem{2014ApJ...789..132R}
    \bibinfo{author}{{Rempel}, M.}
    \newblock \bibinfo{title}{{Numerical Simulations of Quiet Sun Magnetism: On the
      Contribution from a Small-scale Dynamo}}.
    \newblock \emph{\bibinfo{journal}{\apj}} \textbf{\bibinfo{volume}{789}},
      \bibinfo{pages}{132} (\bibinfo{year}{2014}).
    \newblock 
    
    \end{thebibliography}

\section*{Acknowledgments}
We appreciate T. Yokoyama, R. Shimada and T. Hanawa for insightful comments on the manuscript.
The results were obtained using the Supercomputer Fugaku provided by the RIKEN Center for Computational Science, the Supercomputer Flow at Nagoya University, and the Cray XC50 provided by the Center for Computational Astrophysics, National Astronomical Observatory of Japan. Funding: This work was supported by MEXT/JSPS KAKENHI (grant no. JP20K14510 (PI: H. Hotta), JP21H04492 (PI: K. Kusano), JP21H01124 (PI: T. Yokoyama), JP21H04497 (PI: H. Miayahara)) and MEXT as a Program for Promoting Researches on the Supercomputer Fugaku (Toward a unified view of the universe: from large-scale structures to planets, grant no. 20351188 (PI: J. Makino)).

\section*{Authors contributions}

H.H. contributed to the design of the project, developed the numerical code, carried out simulations, performed analysis, and wrote the first draft of the paper. K.K. contributed to the design of the project, interpretation of the result, and writing of the final draft.

\section*{Competing interests}

The authors declare no competing interest.

\section*{Methods}

\section*{Data availability}

We have opted not to make R2D2 code publicly available. Running R2D2 code requires export assistance and appropriate computer system. The numerical method is explained in our previous publication in detail\cite{2014ApJ...786...24H,2015ApJ...798...51H}. 
The data generated, analysed, and presented in this study are available at \url{https://doi.org/10.5281/zenodo.5003258}.
\section*{Numerical simulation}
We solve the three-dimensional magnetohydrodynamic equations in spherical geometry $(r,\theta,\phi)$ with an extended version of the R2D2 code\cite{2019SciA....eaau2307,2020MNRAS.494.2523H}. 
The equations for the calculation are:
\begin{align}
\frac{\partial \rho_1}{\partial t} =& -\frac{1}{\xi^2}\nabla\cdot\left(\rho \bm{v}\right), \\
\frac{\partial}{\partial t} \left(\rho \bm{v}\right) =& -\nabla\cdot\left(\rho \bm{vv}\right) - \nabla p_1 - \rho_1 g \bm{e}_r
+ \frac{1}{4\pi}\left(\nabla\times\bm{B}\right)\times\bm{B} + 2\rho \bm{v}\times\bm{\Omega}_0 \\
\frac{\partial \bm{B}}{\partial t} =& \nabla\times\left(\bm{v}\times\bm{B}\right), \\
\rho T\frac{\partial s_1}{\partial t} =& -\rho T \left(\bm{v}\cdot\nabla\right)s + Q_\mathrm{rad}, \\
p_1 =& \left(\frac{\partial p}{\partial \rho}\right)_s \rho_1 + \left(\frac{\partial p}{\partial s}\right)_\rho s_1,&
\end{align}
where $\rho$, $\bm{v}$, $\bm{B}$, $s$, $T$, $g$, $\bm{\Omega}_0$, and $Q_\mathrm{rad}$ are the density, the fluid velocity, the magnetic field, the specific entropy, the temperature, the gravitational acceleration, the system rotation, and the raidative heating. $\xi$ is the factor from the reduced speed of sound technique\cite{2012A&A...539A..30H}. The subscript 1 indicates the perturbation form the zeoth order spherically symmetric background from the Model S\cite{1996Sci...272.1286C}.
For the detailed method, the readers can find information in our previous publications\cite{2014ApJ...786...24H,2015ApJ...798...51H}. In this study, the calculation domain extends from $0.71R_\odot$ to $0.96R_\odot$ in the radial direction. The whole sphere is covered with the Yin-Yang grid \cite{2004GGG.....5.9005K}. The equations are solved with the fourth-order space-centred method and four-step Runge--Kutta method for time integration. We do not include any explicit diffusivity, and only the artificial viscosity with slope limitter\cite{2014ApJ...789..132R} is used. Non-penetrate and stress-free boundary conditions are used for the top and bottom boundaries. The horizontal and vertical magnetic field boundary conditions are used at the bottom and top boundaries, respectively. \par
We average quantities in the period from 3600 to 4000 days to show the results.

\section*{Normalization of energy spectra}
We adopt a standard way to normalize the energy spectra adopted in our field. When the RMS velocity $v_\mathrm{RMS}(r)$ and magnetic field $B_\mathrm{RMS}(r)$ are defined, the kinetic $E_\mathrm{k}$ and magnetic $E_\mathrm{m}$ energy spectra are normalized to satisfy the relation:
\begin{align}
    \frac{1}{2}\rho_0 v_\mathrm{RMS}^2 &= \sum_{\ell>0}\frac{E_\mathrm{k}(\ell)}{r}, \\
    \frac{B_\mathrm{RMS}^2}{8\pi} &= \sum_{\ell>0}\frac{E_\mathrm{m}(\ell)}{r}.
\end{align}

\end{document}